# Synthesizing like a chemist: an iterative, feedback-driven loop for materials discovery

*Fang Sheng[1], Steven B. Torrisi[2], Amanda Volk[2], Kevin Tran[2], Koki Nakano[2], Brian W. Anthony[1], *Tonio Buonassisi[1]

[1]Department of Mechanical Engineering, Massachusetts Institute of Technology, 77 Massachusetts Ave, Cambridge, Massachusetts, 02139, United States

[2]Toyota Research Institute, 4440 El Camino Real, Los Altos, California, 94022, United States

*Correspondence to Fang Sheng (shengf1227@gmail.com); Tonio Buonassisi (buonassi@mit.edu)

**Abstract**

Most computationally predicted materials are never synthesized because conventional synthesis optimization is slow, expertise-dependent, and iterative. Here we present a closed-loop framework that automates this expert workflow by placing human tacit knowledge in the loop through a large language model (LLM) that distills synthesis knowledge from the literature, high-throughput hyperspectral imaging for rapid film evaluation, and multi-objective Bayesian optimization guided by experimental feedback. In a paired optimization campaign, LLM-assisted initialization produced more Pareto-optimal samples and higher hypervolume than a Latin hypercube sampling baseline at matched trial counts, and this advantage persisted throughout iterative optimization. We demonstrate the framework by synthesizing the previously unreported perovskite-inspired compound $Rb_3BiI_6$ as thin films and validating the optimized films by optical bandgap analysis and X-ray diffraction. The framework transforms synthesis prediction from single-shot recommendation to iterative learning, providing a generalizable strategy to accelerate automated and fully autonomous experimental materials discovery.

# Introduction

Accelerating the pace of materials discovery ultimately depends on solving the synthesis challenge.[1,2] Over the past decade, machine learning (ML) and generative models have dramatically expanded the number of computationally proposed compounds[3–5] for catalysis [6], electrochemistry[7,8], and energy storage and conversion[9,10]. However, the ability to synthesize those candidates has not kept pace with the rate at which they can be proposed.[11,12] A central reason lies in how synthesis is addressed. Many automated and data-driven synthesis-planning approaches treat synthesis as a static "one-shot" mapping from a target material to a recipe, where a single set of precursors and processing conditions is predicted and evaluated once on the basis of previously reported data.[13,14] Iterative, feedback-driven optimization is well established in reaction chemistry — Bayesian reaction optimization is explicitly sequential[15]— but recipe prediction for inorganic materials has largely remained a single-pass mapping. Expert human chemists work very differently. They instead rely on an iterative process that keeps human tacit knowledge in the loop: they formulate an initial hypothesis from prior knowledge and analogy, proceed iteratively, interpret the outcome of each attempt, and adjust the synthesis conditions accordingly.[15–17] The effectiveness of this workflow relies on tacit, difficult-to-formalize knowledge accumulated through years of hands-on experimentation. Yet unsuccessful experiments, intermediate optimization trajectories, and the reasoning behind individual adjustments are rarely reported and documented in the scientific literature.[18–20] The resulting scarcity of training data makes it difficult for artificial intelligence (AI) systems to learn, let alone reproduce, the iterative and intuition-guided strategies that make human chemists effective.[21] This limitation helps explain why the synthesis bottleneck persists even as predictive capabilities continue to improve.[22]

To understand how this bottleneck manifests in practice, we examined how human experts approach the synthesis of a novel material in our own laboratory by reviewing experimental records and logs compiled by previous lab members during the synthesis of $Cs_2AgSbBr_6$ (**Supplementary Fig.1**). We observed that, once a target composition is selected, researchers iteratively adjust synthesis conditions based on domain expertise, visual observations, and instrumental feedback until the desired film quality is achieved. This process is inherently feedback-driven, but it remains slow and dependent on individual experience. Importantly, we found that each step of this expert workflow, from knowledge-informed proposal through rapid feedback to iterative refinement, can be translated into a scalable and quantitative form (**Supplementary Fig.2**), suggesting that materials synthesis can be automated and accelerated through the integration of ML and AI.

Motivated by this observation, we developed a closed-loop framework that mirrors and automates the expert synthesis workflow (**Fig.1**). The framework consists of three major components. First, the domain knowledge traditionally supplied by a researcher's background and experience is extracted using a large language model (LLM), which distills relevant synthesis information from the literature and proposes informed initial synthesis conditions. Second, slow and manual film characterization is replaced by three high-throughput hyperspectral-imaging-based metrics that quantify film coverage, uniformity, and a proxy for phase purity. Third, the iterative decision-making process is implemented through multi-objective Bayesian optimization (MOBO), which uses accumulated hyperspectral feedback to propose synthesis conditions across successive optimization rounds.

This closed-loop workflow improves experimental efficiency by reducing the number of required synthesis trials through LLM-guided domain knowledge while accelerating experimental feedback through rapid hyperspectral characterization. Compared with conventional Latin hypercube sampling (LHS) initialization, LLM-assisted initialization provides a more effective warm start for the optimization campaign by generating higher-quality initial samples. In a single characterization workflow, approximately 30,000 spectra can be processed to extract bandgaps and perform further distribution analysis within minutes. Over the full optimization campaign, our framework reached higher film-quality objective values than the baseline at the same total number of experimental trials. We applied this framework to the synthesis of the perovskite-inspired composition $Rb_3BiI_6$, for which we found no prior experimental report, and characterized the optimized films by optical bandgap analysis, X-ray diffraction (XRD) and density-functional-theory-based structural relaxation, which are respectively consistent with formation of the target phase and with its structural plausibility. By embedding human tacit synthesis knowledge into an autonomous iterative learning loop that comprises of LLM knowledge extraction, high-throughput characterization, and Bayesian optimization, this work demonstrates a generalizable strategy for bridging the gap between computational materials prediction and experimental realization by moving materials synthesis from one-shot prediction toward feedback-driven optimization, thereby accelerating the discovery and validation of new inorganic materials.

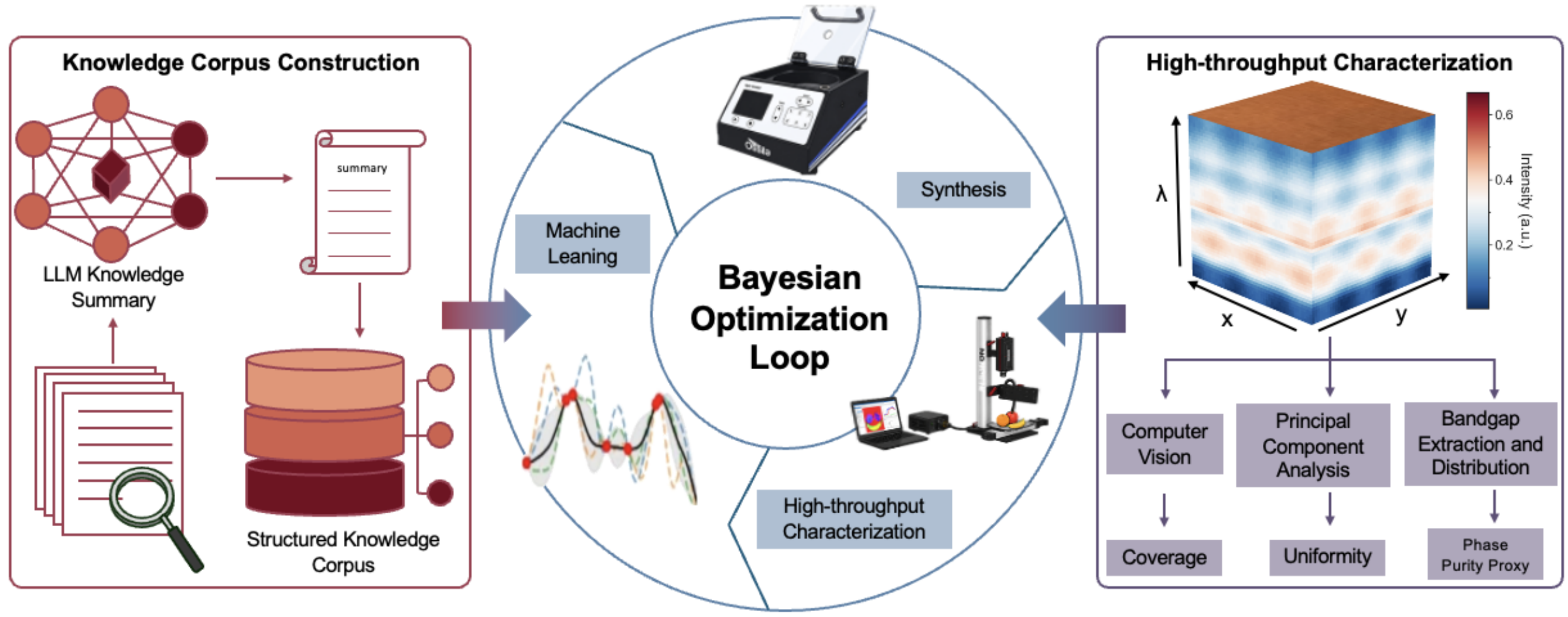

**Fig 1. Diagram of the overall framework for automating and accelerating novel materials synthesis.**

# Results

## High-throughput Characterization using Hyperspectral Imaging

High-throughput characterization is critical for accelerating the acquisition of experimental feedback and, in turn, the optimization process.[23,24] For spin-coated films, we focused on three key metrics to evaluate synthesis quality: coverage, uniformity, and phase purity. Traditionally, coverage is normally assessed by manual visual inspection. Uniformity is measured by a profilometer or atomic force microscope (AFM).[25] Phase purity is determined via X-ray diffraction (XRD) by matching peaks to reference patterns.[26] These methods are largely qualitative or labor-intensive, and require sequential, single-sample measurements, which makes them impractical for the throughput demanded by automated synthesis optimization campaigns involving large sample numbers and quick learning cycles.[27,28]

To address the issue, we developed a pipeline that extracts all three metrics directly from a single datacube obtained from hyperspectral imaging measurement (**Fig.2a**). Hyperspectral imaging can provide both spatial and spectral information on multiple films by performing line-scanning in several minutes. Leveraging diverse data analysis algorithms, this unified approach makes full use of encoded information in the datacube and eliminates the need for separate characterization instruments and sample handling steps. For coverage, a computer vision-based segmentation algorithm distinguishes film-covered regions from the glass substrate (**Fig.2b**). The coverage is quantified as the fraction of film-covered pixels relative to the total pixels of the substrate area. Image-derived coverage and defect metrics have previously been used to guide the optimization of solution-processed perovskite films.[29] Uniformity is evaluated by examining spectral consistency across all pixels (**Fig.2c**). We performed principal component analysis (PCA) on all pixel spectra and generated a detrended PC1 score map as shown in **Fig.2d**, from which uniformity is defined as the average roughness of this 2D score surface. This treatment assumes that the spatial variation of interest is captured by PC1; heterogeneity expressed predominantly in higher components would not be registered by the metric. While optical measurements do not directly probe structural or compositional properties, they can serve as a rapid proxy for identifying the presence of multiple phases through material-specific characteristics such as bandgap.[30,31] We applied the Kubelka–Munk function to convert diffuse reflectance spectra into Tauc plots[32–34] and developed an automated algorithm that identifies the positive region of the first derivative of each Tauc plot and performs linear fitting to extract the bandgap (**Fig.2e**). A single Tauc exponent is applied to every pixel, so the fitted quantity should be read as an effective optical edge rather than as a transition-resolved bandgap[33], and the per-pixel precision of automated edge fitting sets the finest bandgap difference the downstream proxy can resolve[34]. By computing bandgaps for every pixel within a film and analyzing the resulting bandgap distribution (**Fig.2f**), we assign a phase purity proxy score based on the number, shape, and position of distribution peaks (see **Methods** and **Supplementary Note. 1** for details). This pipeline enables scalable and quantitative assessment of all three film quality metrics from a single measurement, capable of extracting approximately 50,000 pixel-level bandgaps in under 5 minutes and delivering coverage, uniformity, and phase-purity scores for an individual film within 10 minutes. This improvement in characterization throughput is what makes the rapid feedback loop required for closed-loop synthesis optimization practical.

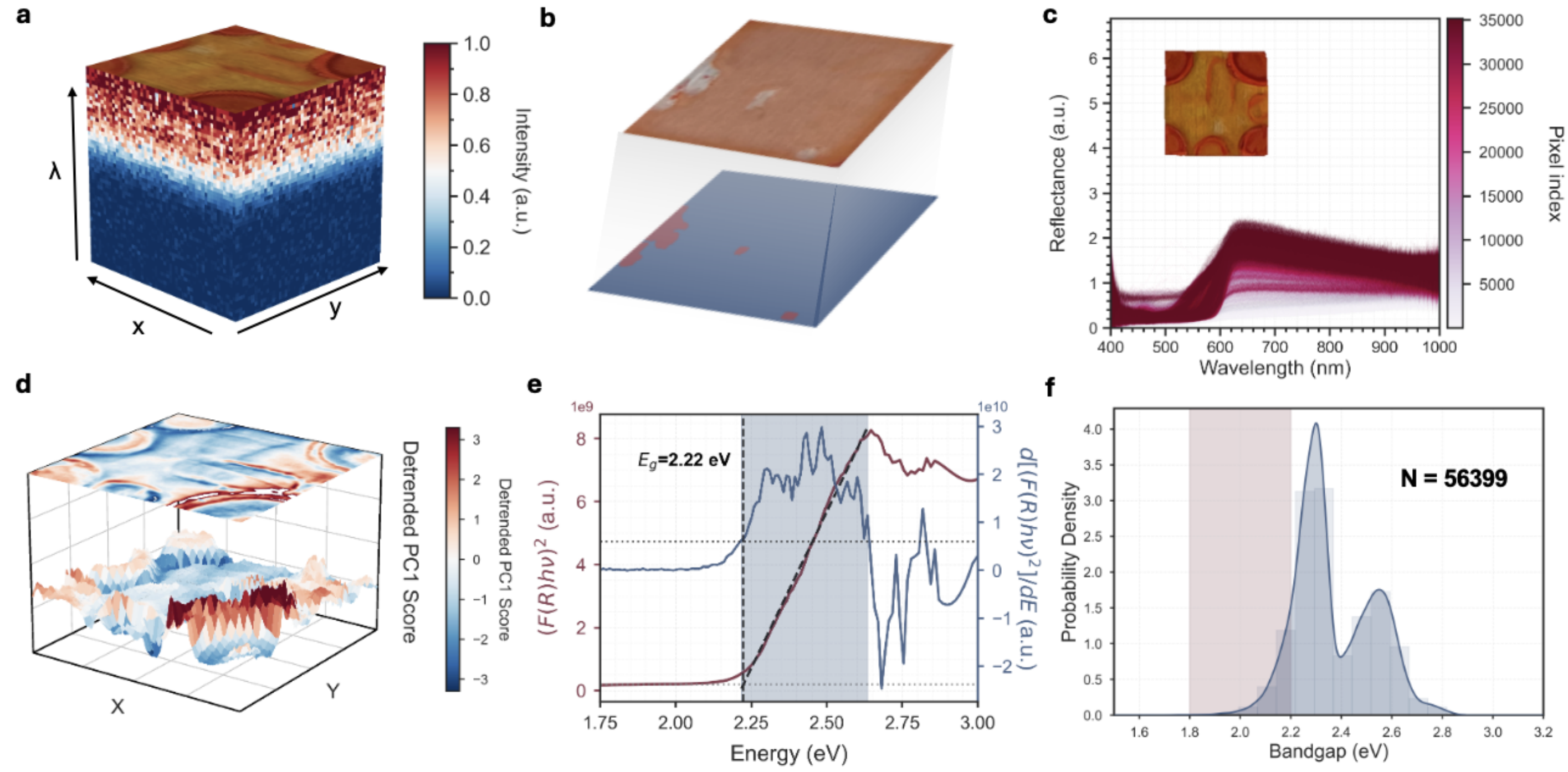

**Fig 2. High-throughput hyperspectral imaging characterization pipeline for film quality. a** Representative hyperspectral datacube containing spatial and spectral information. **b** Schematic illustration of film coverage estimation using computer-vision-based segmentation. **c** Diffuse reflectance spectra extracted from all pixels within the film region. **d** Representative PC1 score map used to quantify film uniformity. **e** Bandgap extraction by identifying the positive region in the first derivative spectrum and fitting the corresponding Tauc plot. **f** Bandgap distribution used to calculate the phase-purity proxy. The red area represents the bandgap region of potential impurities.

## LLM-assisted initialization strategy

In conventional materials discovery workflows, researchers typically initiate the first synthesis trial based on prior experience and intuition. With the advent of machine learning and data-driven approaches, Bayesian optimization (BO) has emerged as a powerful tool for navigating high-dimensional parameter spaces,[35] but it usually initializes this search through space-filling sampling.[36] Though this approach is unbiased, it fails to leverage the domain expertise that guides human researchers[37,38] and can be inefficient when the accessible parameter space is large relative to the experimental cost.[39,40] In this work, we automate the incorporation of human domain knowledge by employing large language models (LLMs) to distill relevant synthesis insights from the literature and inject them into the BO initialization, enabling a more efficient starting point for the optimization loop. Large language models have previously been used to warm-start Bayesian optimization in this way, both zero-shot[41] and adaptively within a ten-dimensional experimental chemistry campaign[42], and other studies report that the benefit is contingent on domain-specific pretraining[43] or find LLM experimental-design agents insensitive to permuted outcome labels[44]. The contribution here is therefore not the use of an LLM prior as such, but a fully offline, temperature-0 extraction stack whose intermediate corpus is auditable, and the use of mined literature to set the search-space bounds rather than only the initial points.

We constructed a workflow that distills human domain knowledge on the synthesis of the target material by retrieving relevant literature and using an LLM to summarize the findings into a structured synthesis knowledge corpus (see more details in **Methods** and **Supplementary Note. 2**). This corpus was then used by the LLM to propose both the relevant parameter space and a set of initialization conditions for multi-objective Bayesian optimization.

We compared this LLM-assisted initialization (LLM-assisted) against the classical and widely-used Latin hypercube sampling (LHS) strategy. LHS was chosen as a space-filling initialization baseline because it selects initial points in multidimensional design space with marginal coverage of each synthesis parameter while avoiding the use of prior knowledge.[45] Other space-filling initialization strategies such as Sobol low-discrepancy sequences, maximin/minimax Latin hypercubes, and sphere packing could also help with warm-start initialization.[46–48] However, LHS was used in this work to specifically evaluate whether literature-derived synthesis knowledge can improve upon a non-informative initialization. One consequence of this design should be stated at the outset: the LLM proposed both the parameter bounds and the initial conditions, whereas the LHS baseline sampled the wider generic spin-coating ranges given in Methods. The two arms therefore differ in the volume of the design space as well as in the sampling rule, and the comparison below measures the combined effect. As shown in **Fig.3a**, the distributions of the five processing parameters under LLM-assisted initialization are tighter than those under LHS, reflecting the LLM's identification of narrower parameter ranges associated with potentially higher-quality films based on extracted domain knowledge. To assess whether this translates into a genuinely better starting point for optimization, we measured coverage, uniformity, and phase purity proxy for all films in the initialization batch and compared the resulting distributions between the two groups (**Fig.3b** and **3c**, and **Supplementary Fig. 3**). Coverage was largely insensitive to the five processing parameters, and both groups achieved uniformly high average coverage. In contrast, LLM-assisted initialization showed clear improvements in the other two metrics, yielding a lower mean and median uniformity score, indicating more uniform films and a higher mean and median phase-purity proxy. Mann–Whitney U tests over the pooled films gave p-values of 0.0013 and 0.0003 for phase-purity proxy and uniformity score, respectively; because each of the 12 conditions per arm was coated in triplicate, these tests treat replicate films as independent observations (see Methods). **Fig.3d** visualizes the full sample population in objective space, with Pareto-front (PF) datapoints marked with outlines. Notably, 9 of the 36 LLM-assisted samples lie on the Pareto front compared with only 3 of 36 for LHS, a three-fold difference in the yield of Pareto-optimal films from initialization alone. This advantage is also supported by the hypervolume trajectories in **Fig.3e**. Here, the hypervolume refers to the objective-space volume dominated by the current Pareto-optimal samples relative to a fixed reference point. LLM-assisted consistently achieved higher hypervolume at matched sample numbers than LHS, suggesting that the LLM-derived prior domain knowledge corpus helps select initial conditions that are not only higher-performing individually but also more informative for mapping the Pareto front.

To quantify this advantage, we defined two complementary metrics.42 One is acceleration factor, capturing how many additional samples LHS requires to reach the same hypervolume achieved by LLM-assisted. The other is enhancement factor, reflecting the relative hypervolume improvement of LLM-assisted over LHS at matched sample numbers. Both metrics are known to depend strongly on the chosen performance threshold and on campaign size, and reported enhancement factors across the self-driving-laboratory literature vary by more than two orders of magnitude[49]; the values below should therefore be read as descriptive of this campaign rather than as transferable figures of merit. As shown in **Fig.3f**, LLM-assisted initialization displays its largest gains in the earliest stages of optimization, where the LLM-assisted approach reaches hypervolume levels within the first few samples that LHS requires substantially more samples to attain. Both factors decay toward a constant value as the two approaches reach their largest hypervolume in the initialization batch. But LLM-assisted consistently maintains a higher hypervolume. Together, these results indicate that, in this campaign, the LLM-distilled domain-knowledge corpus provided a more data-efficient starting point than space-filling initialization.

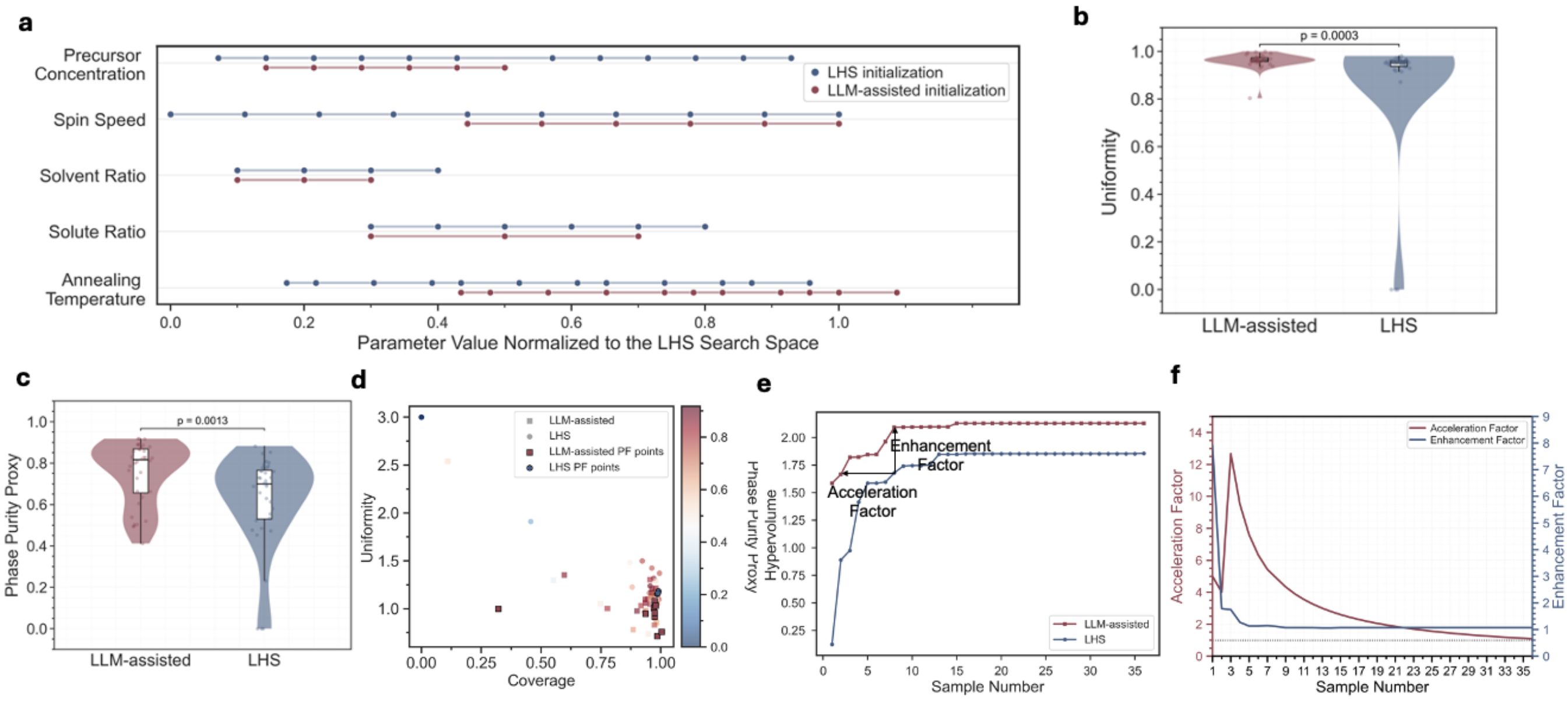

**Fig 3. Comparison between LLM-assisted MOBO initialization (LLM-assisted) and conventional Latin hypercube sampling (LHS). a** Comparison of the initial synthesis conditions proposed by LLM-assisted and LHS, shown as normalized parameter values for spin speed, precursor concentration, annealing temperature, solvent ratio, and solute ratio. **b** Violin plots of the distributions of the uniformity for films synthesized from LLM-assisted and LHS initial conditions. Uniformity was normalized to [0,1] by $U_{\mathrm{score}}=1-\frac{U-U_{\mathrm{min}}}{U_{\mathrm{max}}-U_{\mathrm{min}}}$. **c** Violin plots of the distributions of the phase purity proxy scores for films synthesized from LLM-assisted and LHS initial conditions. **d** Initialization round results from in objective space with Pareto front (PF) points highlighted. **e** Hypervolume evolution as a function of sample number for LLM-assisted and LHS. **f** Acceleration factor (left axis) and enhancement factor (right axis) of LLM-assisted relative to LHS as a function of sample number.

## Multi-Objective Bayesian Optimization Loop

To demonstrate the framework's ability to automate and accelerate the synthesis of novel inorganic materials, we applied it to the synthesis of the perovskite-inspired composition $Rb_3BiI_6$, for which we found no prior experimental report, using a conventional MOBO workflow as the baseline for comparison. The framework is referred to as LLM-BO, whereas the baseline is referred to as regular BO. Except for the initialization stage, all subsequent experimental rounds follow the same MOBO algorithm. We defined a five-dimensional parameter space spanning the key processing conditions: spin speed, precursor concentration, annealing temperature, solvent ratio (DMF:DMSO) and solute ratio ($RbI$:$BiI_3$). The three objectives are coverage, uniformity, and phase-purity proxy, corresponding to the metrics accessible through our high-throughput characterization pipeline. Each workflow consisted of one initialization round followed by three optimization rounds. Beyond initialization, both workflows followed an identical batch MOBO procedure. At each round, accumulated experimental observations were used to update a surrogate Gaussian process (GP) model, and new candidate conditions were proposed via the noisy expected hypervolume improvement (qNEHVI) acquisition function.[50]

**Fig. 4a** visualizes the evolution of the sampled conditions in a low-dimensional embedding of the parameter space across both initialization and optimization rounds. During the initialization stage, the LLM-BO samples occupy a more compact region than those generated by the regular BO baseline, reflecting the narrower, knowledge-informed search space proposed by the LLM. In contrast, regular BO begins with broader space-filling sampling across the design space. As optimization proceeds, the LLM-BO samples progressively converge toward a localized region of parameter space, whereas the regular BO samples remain more broadly distributed. Consistent trends are observed in the round-by-round processing parameter distributions and film appearances (**Supplementary Fig.4**), where LLM-BO exhibits a narrower initialization distribution and more rapid convergence toward high-quality films with visually improved coverage and uniformity.

The advantage provided by the warm initialization is quantified by the hypervolume trajectories in **Figure 4b**, where the LLM-BO begins with a substantially higher hypervolume than the regular BO baseline. Importantly, this advantage is not limited to the initialization stage. Although both workflows improve as optimization proceeds, LLM-BO consistently maintains and further extends its hypervolume advantage throughout the campaign. This result demonstrates that the framework not only benefits from a superior initialization but also preserves that advantage during sequential learning, ultimately achieving a higher final hypervolume for the same number of experimental trials. Because each workflow was run once, with a single random seed, the magnitude of this advantage cannot be separated from run-to-run variability.

**Figure 4c** shows the overall objective performance across the optimization campaign, using the sum of all three objectives as a single performance metric. Since the uniformity metric is originally positive-valued and lower-is-better, it was normalized to the [0, 1] range (with higher values indicating better performance) to match the scales of coverage and phase purity proxy (see **Methods** for details).

The LLM-BO workflow identifies high-performing samples early in the campaign and continues to propose competitive candidates in later rounds, whereas the baseline BO workflow generates a larger proportion of low-performing samples during early rounds. The dashed lines mark the best cumulative objective value achieved by each workflow as a function of sample number. While both approaches eventually converge on strong candidates, the LLM-BO workflow reaches high-performing regions of the design space with fewer low-value trials, underscoring its advantage in experimental efficiency, and ultimately attains a higher final best-sample performance than the baseline. Consistent results are also observed when objective uniformity and phase purity proxy are analyzed individually (**Supplementary Fig. 5**). This conclusion is also consistent with the spatial clustering in the initialization round and at the end of the optimization when the pooled experimental data are visualized in pairwise parameter space, where LLM-BO samples concentrate near the predicted optimum of a shared summary surrogate while regular BO samples are more broadly distributed across the design space (**Supplementary Fig. 6**).

To illustrate the performance advantage of the LLM-BO workflow more directly, we defined a "good sample" as one satisfying coverage > 0.90, uniformity < 1.0, and phase-purity proxy > 0.90, and computed the success rate as the ratio of the number of samples meeting this criterion to the total number of samples. As shown in **Figure 4d**, LLM-BO exhibits a continuously increasing success rate and achieves the highest final value of any round across either workflow. The conventional BO baseline, by contrast, produced no additional good samples in the final round, resulting in a slight decrease in its success rate. This suggests that the benefit of LLM-assisted initialization can extend beyond improving early hypervolume by increasing the likelihood of identifying experimentally successful conditions throughout the entire optimization campaign.

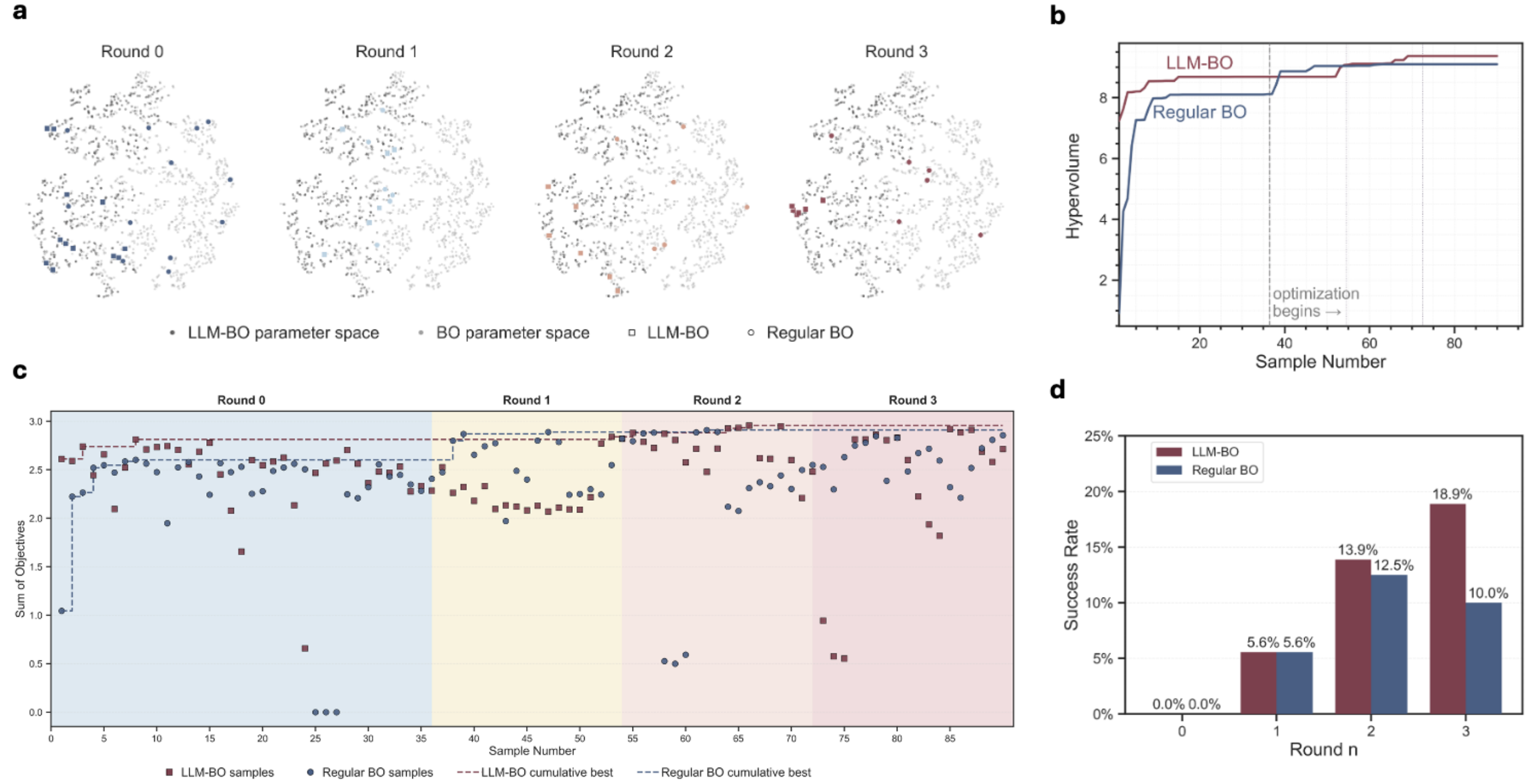

**Fig 4. Comparison between LLM-BO MOBO and conventional MOBO throughout the optimization campaign. a** t-SNE visualization of the parameter space for $Rb_3BiI_6$ film synthesis during the optimization campaign. The design space for regular BO is colored in light gray and that for LLM-BO is colored in dark grey. The synthesis conditions for regular BO are represented by a circle while those for LLM-BO are represented by a square. **b** Hypervolume evolution of LLM-BO and BO over the optimization as a function of sample number. **c** Sum of objective values for every synthesized sample throughout the optimization campaign. Samples are colored by round numbers. LLM-BO samples are represented by squares and BO samples are represented by circles. The dashed lines indicate the best-performing sample achieved by LLM-BO and regular BO. **d** Success rate across the optimization campaign for LLM-BO and regular BO.

## Optical and Structural Assessment of $Rb_3BiI_6$ Films

Given that the LLM-BO framework demonstrated better overall performance across all three objectives after three optimization rounds, we used the resulting experimental dataset to fit a final surrogate model (See details of model validation in **Supplementary Fig.7-9**) and identify synthesis conditions lying on the predicted Pareto front. **Figure 5a** and **5b** show the pairwise predicted objective landscape and the corresponding model uncertainty, respectively, with representative images of synthesized films overlaid at selected Pareto-front conditions. See **Supplementary Fig.10** for another two pairs and **Supplementary Fig.11-13** for pairwise visualizations in design space. These films exhibit the high coverage and visual uniformity expected from their predicted objective values, indicating that the surrogate model's predictions translate into experimentally realizable, high-quality samples.

To assess the phase-purity proxy against a structural measurement, we performed hyperspectral imaging on a representative film selected from the Pareto front. The corresponding Tauc plot of the average reflectance spectrum (**Figure 5c**) yields an optical bandgap of $E_g$ = 2.26 eV under an indirect-transition Tauc exponent. Density functional theory (DFT) calculations predict an indirect bandgap of 2.74 eV (**Supplementary Fig. 14**), higher than the experimental value. Several effects contribute to this difference and they act in opposing directions: excitonic absorption, defects and disorder in real films, and the omission of spin–orbit coupling all lower the measured edge relative to a scalar-relativistic calculated gap[51], whereas the self-interaction error of the PBE functional typically underestimates gaps in bismuth halides. The agreement obtained here therefore reflects a partial cancellation of two large opposing errors rather than a converged prediction. To further confirm the phase identity and crystal structure, we performed X-ray diffraction (XRD) on the same film (**Figure 5d**). Pawley refinement of the diffraction pattern is consistent with a cubic cell of lattice parameter a = 12.64 Å, in reasonable agreement with the cell expected for this composition by analogy with the related compound $Cs_3SbBr_6$ documented in the Materials Project (See details in Methods). Because Pawley refinement fits reflection intensities freely and uses no atomic coordinates, it constrains the unit cell and the peak profile rather than the atomic arrangement and provides no route to quantitative phase fractions. The pattern shows no strong reflections beyond those indexed by this cell, with minor peaks attributable to the secondary phase $Rb_3Bi_2I_9$[52,53]. Alternative RbI-rich products, in particular $Rb_7Bi_3I_{16}$, and a second polymorph of $Rb_3Bi_2I_9$ were not explicitly excluded, and the assignment should be regarded as provisional pending Rietveld refinement against an atomistic model and an independent measurement of the Rb:Bi:I stoichiometry. Together, these results indicate that the LLM-BO framework accelerates the identification of high-performing synthesis conditions and yields films whose optical and diffraction signatures are consistent with the target perovskite-inspired composition.

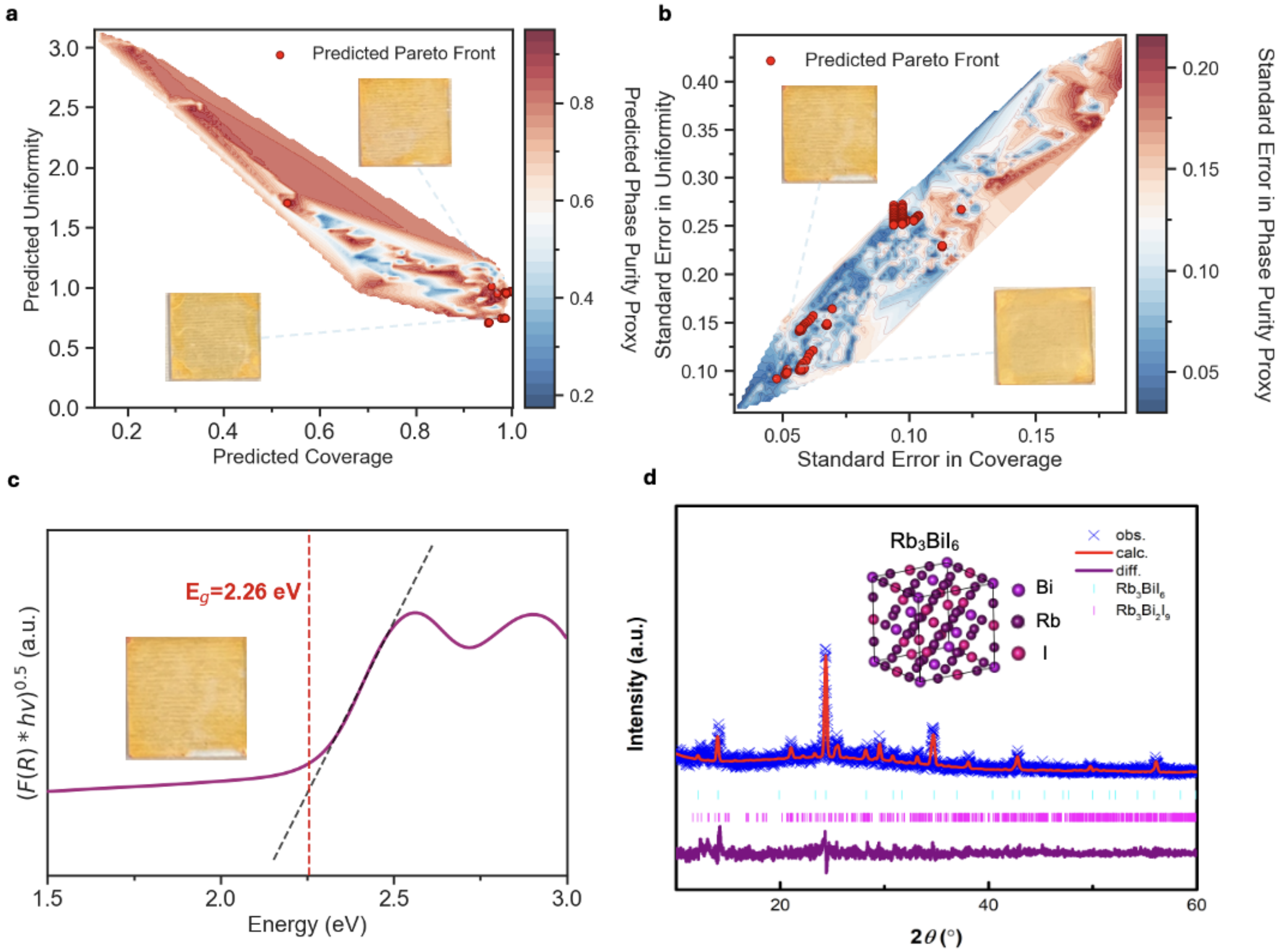

**Fig 5. Validation of $Rb_3BiI_6$ films after optimization with the proposed framework. a** Predicted value landscape in objective space with synthesis conditions on the Pareto front labeled in red points. Representative film images show the corresponding synthesized films to the specific PF conditions. **b** Predicted uncertainty landscape in objective space with synthesis conditions on the Pareto front labeled in red points. Representative film images show the corresponding synthesized films to the specific PF conditions. **c** Indirect Tauc plot and bandgap extraction result of an optimized $Rb_3BiI_6$ film. **d** The Pawley refinement for experimental XRD pattern of optimized $Rb_3BiI_6$ film with cubic structure.

# Discussion

This study presents a multi-objective Bayesian optimization framework that embeds human tacit knowledge in the loop by integrating LLM-assisted knowledge corpus construction and domain-knowledge injection with single-instrument and high-throughput characterization techniques. Rather than treating synthesis as a one-shot prediction problem, the framework captures the way expert chemists iteratively combine prior knowledge with experimental feedback to refine synthesis conditions. Using this framework, we synthesized thin films of a perovskite-inspired composition for which we found no prior experimental report, with high coverage, improved uniformity and improved phase purity proxy scores, and reached those objective values with fewer low-value trials than the baseline workflow.

Three features of the result are worth restating. First, the advantage conferred by LLM-assisted initialization appeared in the first batch and was still present at the end of the campaign, rather than being erased once both workflows switched to the same Bayesian optimization procedure. Second, deriving all three objectives from a single hyperspectral datacube removed the sequential, single-sample characterization steps that would otherwise set the cycle time of the loop, delivering coverage, uniformity and phase-purity scores for a film within ten minutes. Third, that same pipeline turned the analysis of the campaign into an automated step, so the volume of data collected was no longer bounded by the effort required to interpret it.

Several limitations remain in this work. First, we selected a relatively small language model that can be run locally on a personal laptop for knowledge corpus summarization and synthesis condition selection. This choice was made to improve the repeatability and reproducibility of the experiments. As a result, the quality of the model responses and its reasoning capability are more limited than those of larger frontier models. Nevertheless, the results demonstrate the potential of automated domain knowledge incorporation via an LLM to accelerate synthesis optimization. Because the framework was not tested with a second model, we cannot say whether the observed advantage is specific to this model or would survive a change of model, prompt set or literature corpus; reported benefits from LLM priors in optimization have been found to depend on domain-specific pretraining[43], and LLM experimental-design agents have in some settings shown no sensitivity to the experimental outcomes they are given[44]. Second, we used optical proxies derived from hyperspectral imaging for rapid film-quality assessment. These proxies are challenging to calibrate against ground truth measurements, which themselves can be difficult to quantify using conventional characterization methods. For example, Pawley refinement of thin-film XRD patterns cannot reliably provide quantitative phase ratios without Rietveld refinement for powder diffraction data. More broadly, automated whole-pattern refinement has been shown to return confident but incorrect phase assignments at scale in autonomous synthesis campaigns[54], so the structural label used to check an optical proxy carries uncertainty of its own. The proxy also cannot separate two phases whose optical edges differ by less than the per-pixel fitting precision, and the uniformity metric registers only the variation captured by the first principal component. Although the proxy metrics may contain some deviations, and the final optimized film still contains minor impurity phases, they effectively guide the iterative optimization process toward improved film quality. Finally, we validated the framework using only one previously unreported material system and spin-coating approach with limited processing parameters. Broader testing across additional compositions, more complicated parameter spaces and synthesis routes will be needed to further establish its generality. Three aspects of the comparison itself also bound the strength of the conclusions. The LLM set both the search-space bounds and the initial conditions in the LLM-assisted arm, whereas the baseline arm sampled the wider generic spin-coating ranges given in Methods; the comparison therefore measures the combined effect of narrowing the search space and of choosing points within it, and a Latin hypercube design run inside the LLM-proposed bounds would be required to separate the two. Each workflow was also run once, with a single seed and 18 optimization samples across five parameters, so the reported acceleration and enhancement factors cannot be separated from run-to-run variability[49]. And the optimization was carried out inside a fixed parameter space and a single deposition route, which leaves open whether an LLM can contribute to the design of a wholly new experimental procedure rather than to the setting of bounds within an established one; its efficacy is also likely to depend on how much relevant literature exists for the target system and on the model's ability to transfer inferences from related compounds.

Overall, this work provides a practical strategy for addressing the materials synthesis bottleneck in data-driven materials discovery by reducing the effort involved in experiment planning and analysis, making it easier to frame novel material discovery as an iterative optimization problem. By combining LLM-derived domain knowledge, rapid high-throughput characterization, and closed-loop Bayesian optimization, the framework translates an expert-driven experimental workflow into an automated and scalable process. Philosophically, this advances the synthesis challenge from "single-shot" to "iterative," in line with expert experimentalist working style. We note that $Rb_3BiI_6$ served here as a demonstration vehicle for the framework rather than as a candidate photovoltaic absorber: a composition with isolated $[BiI_6]^{3-}$ octahedra is of low electronic dimensionality, a class for which heavy carrier effective masses and deep defect states are expected on structural rather than processing grounds[55], so "perovskite-inspired" should not be read here as implying device promise. More broadly, this approach could be extended to other solution-processed inorganic materials and functional thin-film systems, helping to bridge the gap between computational materials prediction and experimental realization.

# Methods

### *Materials*

N, N-Dimethylformamide (DMF, anhydrous, 99.8%), dimethyl sulfoxide (DMSO, anhydrous, 99.9%) and isopropanol (IPA, anhydrous, 99.9%) were purchased from Sigma-Aldrich. Rubidium iodide (RbI, 99.8%) and bismuth iodide ($BiI_3$, 99.999%) were purchased from Thermo Scientific. All chemicals were used without further purification.

### *Film Synthesis*

Plain glass substrates were used for all film depositions. The cleaning procedure included sonicating under deionized water and IPA for 10 minutes under sonication. Prior to use, the substrates were cleaned by sequential sonication in deionized water and isopropanol (IPA) for 10 min each, then dried. The dried substrates were subsequently treated in an Ossila UV-ozone cleaner (model L2002A3) for 10 min. The $Rb_3BiI_6$ precursor solutions were prepared by dissolving RbI and $BiI_3$ in a mixed solvent of DMF and DMSO. The solute ratio, solvent ratio, and total precursor concentration were varied according to the specific experimental conditions. The solutions were spin-coated onto the glass substrates using an Ossila spin-coater at a certain rotation speed for 30 s. The as-deposited films were then annealed at a specific temperature for 10 min. The spin-coating and annealing process were conducted in fume hood. After annealing, the films were transferred to a desiccator for storage until further characterization. All conditions were repeated three times.

### *LLM Knowledge Extraction Framework*

Relevant literature was first identified using Google Scholar queries executed through SerpAPI. To ensure broad coverage of the target material and related synthesis context, four complementary search queries were used: (1) a primary full-name query, "rubidium bismuth iodide spin coating thin films"; (2) an abbreviation-based query, "Rb3BiI6 spin coating thin films"; (3) a combined formula and full-name query,

"Rb3BiI6 rubidium bismuth iodide film fabrication"; (4) a perovskite-context query, "Rb3BiI6 rubidium bismuth iodide perovskite spin coating." The resulting literature links were collected, deduplicated, and downloaded as PDF files either automatically or semi-automatically, depending on document accessibility, into a designated local folder. The collected PDF folder was then analyzed using a fully local retrieval-augmented LLM workflow based on PaperQA[56] with locally hosted language models accessed through Ollama. Online metadata enrichment was disabled to ensure that all extracted information originated only from the collected documents. The local LLM Qwen2.5-14B[57] was used to extract synthesis-relevant knowledge from the PDF corpus (**Supplementary Note 2** and **3**). To obtain a structured and comprehensive synthesis knowledge base, the extraction prompts were organized into four categories, including direct synthesis information for $Rb_3BiI_6$, synthesis knowledge from related $A_3BiI_6$ compounds, element-based synthesis information involving Rb-, Bi-, and I-containing compounds, and general synthesis notes related to troubleshooting, morphology control, and characterization guidance (**Supplementary Note 2**). The extracted information was consolidated into a structured JSON knowledge base, which was subsequently used to define a chemically reasonable search space for Bayesian optimization. Specifically, Qwen2.5-14B was provided with the extracted synthesis knowledge and prompted to recommend numerical parameter ranges, step sizes, and rationales for each optimization variable. Using this predefined search space, Qwen2.5-14B was then prompted to propose an initialization batch of 12 experimental conditions for Bayesian optimization. The prompt instructed the model to prioritize experimentally feasible conditions, exclude precursor concentrations that were unlikely to yield clear precursor solutions, and balance conservative literature-guided conditions with exploratory conditions that broadened the coverage of the design space. To improve the reproducibility of the LLM-generated responses, the sampling temperature for all prompts was set to 0.

### *High-throughput Characterization*

*Hyperspectral imaging*

Hyperspectral imaging was performed using a Resonon Pika L line-scan camera paired with a Techniquip 21DC PowerLine halogen light source. All samples were measured under ambient air conditions without encapsulation. The camera was configured to collect spectra over a range of 400–1200 nm at 1 nm spectral resolution, with a frame rate of 5 fps and an integration time of 76.5 ms.

*Coverage*

Watershed segmentation was applied to the datacube acquired through hyperspectral imaging to identify the covered film area. The coverage is calculated by dividing the number of film-covered pixels by the total number of pixels in the substrate region.

*Uniformity*

Uniformity was quantified from pixel-resolved hyperspectral diffuse reflectance data. Pixel spectra from the covered area of each film were extracted and smoothed using a Savitzky–Golay filter to suppress high-frequency instrumental noise. Principal component analysis was then applied to the preprocessed spectra. The first principal component score (PC1) was used as a scalar representation of the dominant spectral variation at each pixel. Next, the PC1 score map was detrended by fitting and subtracting a linear spatial plane. This detrending step aims to remove global spatial gradients such as illumination gradients and retain the residual local variation used to evaluate film uniformity. The uniformity score was defined as the mean absolute deviation of the detrended PC1 scores:

$$R_a = \frac{1}{N}\sum_{i=1}^{N}\left|p_i^{det} - \bar{p}^{det}\right|$$

where N is the number of valid pixels and $\bar{p}^{det}$ is the mean detrended PC1 score over those pixels. This quantity is analogous to an average roughness applied to the detrended PC1 score map. Lower $R_a$ indicates smaller pixel-to-pixel spectral variation and therefore higher film uniformity, whereas larger $R_a$ indicates stronger spatial heterogeneity and lower uniformity.

*Phase Purity Proxy*

The phase-purity proxy was calculated from the pixel-wise bandgap distribution extracted from hyperspectral reflectance measurements. For each film, all pixel bandgaps were first collected from the segmented film region and represented as an observed bandgap distribution. A continuous probability density function P(x) was then estimated using Gaussian kernel density estimation (KDE) over a fixed energy grid from 0.8 to 3.2 eV, where x denotes the optical bandgap energy. Potential impurity phases were identified based on the elemental composition of the target material system. Literature-reported bandgap values of these possible impurity phases were used to define undesired bandgap windows. For example, in the $Rb_3BiI_6$ synthesis space, $Rb_3Bi_2I_9$ was considered as a highly possible impurity with an approximate undesired bandgap range of 1.90–2.10 eV.[53] An ideal phase-pure film is expected to exhibit a dominant and localized bandgap feature. Therefore, a narrow Gaussian probe was used to quantify the degree to which the observed bandgap distribution contained a sharp target-like peak outside the predefined impurity windows. For a candidate target bandgap center μ, the target-phase overlap was defined as:

$$G(\mu) = \int P(x)\,\mathcal{N}(x;\mu,\sigma_g^2)\,dx$$

where $\mathcal{N}(x;\mu,\sigma_g^2)$ is a Gaussian probe centered at μ, and $\sigma_g$ controls the width of the target-phase probe. Because the true target bandgap was not assumed a priori, the optimal target peak position was determined by a dense grid search over the allowed bandgap range, excluding expanded impurity windows $[\mu_{j,min}-\delta, \mu_{j,max}+\delta]$ as below:

$$\mu^* = \arg\max_{\mu \in \mathcal{I}_{\text{allowed}}} G(\mu)$$

where $\mathcal{I}_{\text{allowed}}$ denotes the bandgap search range. δ is a tolerance margin used to improve robustness against uncertainty in literature-reported impurity bandgaps and multimodal bandgap distributions.

To penalize impurity contributions, the overlap between the observed bandgap distribution P(x) and each undesired impurity bandgap window was calculated using a Gaussian impurity probe:

$$B_j = \int P(x)\,\mathcal{N}(x; \mu_j, \sigma_{\text{bad}}^2)\,dx$$

where $\mu_j$ is the center of the impurity bandgap region and $\sigma_{bad}$ controls the width of the impurity penalty. If multiple impurity regions were considered, the total impurity penalty was calculated as the average overlap:

$$B = \frac{1}{m} \sum_{j=1}^{m} B_j$$

where m is the number of impurity features.

To account for pixels without absorption features within the measured spectral range, a valid-pixel fraction was also included:

$$f_{\text{valid}} = \frac{N_{\text{valid}}}{N_{\text{total}}}$$

where $N_{valid}$ is the number of pixels for which a bandgap could be fitted and $N_{total}$ is the total number of pixels in the segmented film region. This term penalizes optically invalid or poorly fitted film regions. The final phase-purity proxy was calculated as

$$S_{\text{phasepurityproxy}} = f_{\text{valid}} \frac{G(\mu^*)}{G(\mu^*) + \lambda B}$$

where $\lambda$ is a weighting factor controlling the strength of the impurity penalty. Unless otherwise stated, $\lambda = 1$ was used. The resulting score ranges from 0 to 1, with higher values indicating a larger fraction of valid pixels, a stronger dominant bandgap feature outside the impurity regions, and reduced overlap with known impurity-associated bandgap signatures.

## *Multi-Objective Bayesian Optimization*

The multi-objective Bayesian optimization code was adapted from https://github.com/PV-Lab/MOBO-Kit. Five experimentally controllable processing parameters were optimized: spin speed, precursor concentration, annealing temperature, solvent ratio (DMF:DMSO), and solute ratio ($RbI:BiI_3$). For the baseline MOBO workflow, the search space was defined to cover commonly used spin-coating synthesis ranges, including spin speeds from 500 to 5000 rpm, precursor concentrations from 0.1 to 1.5 mol/L, annealing temperatures from 100 to 300 °C, solvent ratios from 1:1 to 4:1, and solute ratios from 2.8:1 to 3.2:1. The search space for the LLM-BO workflow was proposed by the local LLM and differs from the baseline ranges above; The search space was discretized according to step size. For model training and acquisition-function optimization, all input parameters were normalized to the range [0, 1]. Suggested continuous candidates were subsequently mapped back to the nearest experimentally allowed discrete values before synthesis. Each experimental condition was repeated three times. Three objectives were surface coverage, uniformity, and phase-purity proxy. Coverage and phase-purity proxy were maximized, whereas the uniformity score was minimized. To formulate the optimization as a maximization problem for all objectives, the uniformity score was multiplied by $-1$ during MOBO model training and Pareto-front analysis. Replicate experimental measurements were treated as individual observations during surrogate-model fitting. For each optimization round, independent Gaussian process surrogate models were fitted for the three objectives using BoTorch.[58] A separate single-task Gaussian process was trained for each objective, and the three models were combined into a model list for multi-objective acquisition-function evaluation. The Gaussian processes used Matérn kernels with automatic relevance determination, Gaussian likelihoods, and standardized outcome transformations. Model predictions were evaluated using posterior mean and uncertainty estimates for each objective. New candidate experiments were selected using the noisy expected hypervolume improvement (qNEHVI) acquisition function[50], which prioritizes conditions expected to improve the Pareto front while accounting for experimental noise and surrogate-model uncertainty. A batch size of six candidate conditions was used for each optimization round. The reference point for hypervolume calculations was defined adaptively as the component-wise minimum of the observed objective values minus 0.01 in the maximization-transformed objective space, [coverage, −uniformity, phase purity proxy]. The qNEHVI acquisition function was evaluated using 32 Sobol quasi-Monte Carlo samples, and acquisition-function optimization used 10 restarts and 256 raw samples.

### *X-ray Diffraction (XRD) and Pawley Refinement*

X-ray diffraction (XRD) measurements were performed using a Rigaku SmartLab X-ray diffractometer with a Cu-Kα source. Grazing-incidence XRD patterns were collected at an incidence angle of 0.5°. The 2θ angle range was 10-60°, with a step size of 0.02° and a scanning speed of 6°/min. The Pawley refinement was performed in GSAS-II using the simulated $Rb_3BiI_6$ crystal structure obtained after structural relaxation. Because Pawley refinement treats reflection intensities as free parameters and does not use atomic coordinates, it constrains the unit cell and the peak profile but does not test the atomic arrangement and yields no route to quantitative phase fractions. No quantitative phase analysis was attempted, and the amorphous fraction was not determined.

### *Calculation of Acceleration Factor and Enhancement Factor*

The acceleration factor was defined as the relative number of samples required by the baseline BO workflow to reach the same hypervolume achieved by the LLM-assisted BO workflow after n samples. The baseline BO workflow was initialized by Latin hypercube sampling (LHS). If the LHS initialization alone did not reach the target hypervolume, samples from subsequent baseline BO optimization rounds were included in the baseline trajectory. For each sample number n in the LLM-assisted BO trajectory, the corresponding hypervolume $HV_{LLM\text{-}BO}(n)$ was used as the target value. The baseline BO trajectory was then searched to identify the smallest sample number k for which $HV_{\mathrm{LHS}}(k) \geq HV_{\mathrm{LLM-BO}}(n)$. The acceleration factor was calculated as:

$$AF(n) = \frac{k}{n}$$

The enhancement factor was calculated by directly comparing the hypervolumes achieved by the two approaches at the same sample number. For each shared sample number n, the enhancement factor was defined as:

$$EF(n) = \frac{HV_{\mathrm{LLM-BO}}(n)}{HV_{\mathrm{LHS}}(n)}$$

### *Calculation of Sum of Objectives*

The sum of objectives was used as a single scalar metric to compare overall sample performance. Surface coverage and phase-purity proxy were already defined on a larger-is-better scale, with values ranging from 0 to 1. In contrast, the raw uniformity score was defined on a lower-is-better scale and had a non-negative, unbounded range. Therefore, the uniformity score was first normalized to the range [0, 1] and transformed into a larger-is-better metric before aggregation. To ensure a consistent normalization across all optimization rounds and methods, the 5th and 95th percentiles of the raw uniformity scores were calculated globally using all samples from both LLM-assisted BO and baseline BO. For each sample, the raw uniformity score (U) was clipped to this global percentile range and then converted to a normalized uniformity score as below:

$$S_{\text{uniformity}} = 1 - \frac{\text{clip}(U, U_5, U_{95}) - U_5}{U_{95} - U_5}$$

where $U_5$ and $U_{95}$ are the global 5th and 95th percentile values of the uniformity scores, respectively. This transformation ensures higher values assigned to more spatially uniform samples while reducing the influence of extreme outliers. The final sum of objectives was calculated as an equal-weighted sum of the three objectives as below:

$$S_{\text{total}} = S_{\text{coverage}} + S_{\text{phasepurityproxy}} + S_{\text{uniformity}}$$

Because each term was scaled to the same range and defined such that larger values indicate better performance, a higher sum of objectives corresponds to better overall film quality.

### *Crystal Structure Relaxation and Theoretical Bandgap Calculation*

First-principles density functional theory (DFT) calculations were performed using the Quantum ESPRESSO package to investigate the crystal structure and electronic structure of $Rb_3BiI_6$. The initial structure was constructed from the crystallographic structure of the chemically similar compound $Cs_3SbBr_6$ (Materials Project ID: mp-1112952) obtained from the Materials Project.[59,60] Cs and Sb atoms were replaced by Rb and Bi, respectively. The lattice parameter was linearly adjusted based on the corresponding ionic radii. The modified crystallographic information file was then converted into Quantum ESPRESSO[61,62] input format. The calculation cell contained 40 atoms. Geometry optimization was first performed to obtain the relaxed atomic configuration. The optimized lattice vectors and fractional atomic coordinates were then used for a self-consistent field calculation. The SCF calculation was carried out using the Perdew–Burke–Ernzerhof (PBE) exchange–correlation functional within the generalized gradient approximation.[63] Plane-wave kinetic energy and charge-density cutoffs were set to 50 Ry and 400 Ry, respectively. The Brillouin zone was sampled using a 4×4×4 Monkhorst–Pack k-point mesh for the SCF calculation. Fixed occupations were used because $Rb_3BiI_6$ is expected to be semiconducting. The electronic convergence threshold was set to $1.0\times10^{-8}$ Ry. Following convergence of the SCF calculation, a band-structure calculation was performed using the self-consistent charge density from the SCF step. Band energies were calculated along the high-symmetry path W→L→Γ→X→W, which is commonly used for cubic perovskite-type structures. The resulting band eigenvalues were post-processed using bands.x to generate plottable band-structure data (Supplementary Fig. 14). Spin–orbit coupling was not included in the present calculation. Because the PBE functional typically underestimates bandgaps in bismuth halides while the omission of spin–orbit coupling raises them, these two errors partially cancel; therefore, the calculated value should not be interpreted as a fully converged prediction[51].

# LLM Usage

Large language models were used in the research portion of this study as documented in the "Methods" and "Supplementary Notes," with the goal of being scientifically reproducible. Separately, frontier LLMs were used as brainstorming aids for message framing, as well as text and figure refinement based on first user drafts.

# Data availability

The source data are available at https://osf.io/cbrk2/overview.

# Code availability

The code is available at https://github.com/PV-Lab/SpinSynth. The multi-objective Bayesian Optimization code is implemented in https://github.com/PV-Lab/MOBO-Kit.

## Acknowledgments

We thank Noor Titan P. Hartono, Janak Thapa and Shijing Sun for providing the experimental log records of $Cs_2AgSbBr_6$ film synthesis optimization. We thank Dr. Basita Das and Alex Love for providing samples used in comparing methods for calculating the phase purity proxy metric. This project was funded by the Toyota Research Institute under the Synthesis Advanced Research Challenge (SARC).

## Author contributions

F.S., S.T., and T.B. conceived this work. F.S. fabricated the samples and collected all the characterization data. F.S. developed all the workflows and algorithms. F.S. performed all the data analysis. S.T., A.V., K.T., B.A. and T.B. discussed the results and supervised the research project. F.S. wrote and all authors edited this manuscript.

## Competing interests

Toyota Research Institute is a wholly owned subsidiary of Toyota Motor Corporation, which is a for-profit entity.